\documentstyle[twoside]{article}
\input psfig.tex
\catcode`\@=11
\long\def\@makefntext#1{
\protect\noindent \hbox to 3.2pt {\hskip-.9pt  
$^{{\eightrm\@thefnmark}}$\hfil}#1\hfill}		%CAN BE USED 

\def\thefootnote{\fnsymbol{footnote}}
\def\@makefnmark{\hbox to 0pt{$^{\@thefnmark}$\hss}}	%ORIGINAL 
	
\def\ps@myheadings{\let\@mkboth\@gobbletwo
\def\@oddhead{\hbox{}
\rightmark\hfil\eightrm\thepage}   
\def\@oddfoot{}\def\@evenhead{\eightrm\thepage\hfil
\leftmark\hbox{}}\def\@evenfoot{}
\def\sectionmark##1{}\def\subsectionmark##1{}}

\oddsidemargin=\evensidemargin
\renewcommand{\thefootnote}{\fnsymbol{footnote}}

\newcounter{sectionc}\newcounter{subsectionc}\newcounter{subsubsectionc}
\renewcommand{\section}[1] {\vspace{12pt}\addtocounter{sectionc}{1} 
\setcounter{subsectionc}{0}\setcounter{subsubsectionc}{0}\noindent 
	{\tenbf\thesectionc. #1}\par\vspace{5pt}}
\renewcommand{\subsection}[1] {\vspace{12pt}\addtocounter{subsectionc}{1} 
	\setcounter{subsubsectionc}{0}\noindent 
	{\bf\thesectionc.\thesubsectionc. {\kern1pt \bfit #1}}\par\vspace{5pt}}
\renewcommand{\subsubsection}[1] {\vspace{12pt}\addtocounter{subsubsectionc}{1}
	\noindent{\tenrm\thesectionc.\thesubsectionc.\thesubsubsectionc.
	{\kern1pt \tenit #1}}\par\vspace{5pt}}
\newcommand{\nonumsection}[1] {\vspace{12pt}\noindent{\tenbf #1}
	\par\vspace{5pt}}

\newcounter{appendixc}
\newcounter{subappendixc}[appendixc]
\newcounter{subsubappendixc}[subappendixc]
\renewcommand{\thesubappendixc}{\Alph{appendixc}.\arabic{subappendixc}}
\renewcommand{\thesubsubappendixc}
	{\Alph{appendixc}.\arabic{subappendixc}.\arabic{subsubappendixc}}

\renewcommand{\appendix}[1] {\vspace{12pt}
        \refstepcounter{appendixc}
        \setcounter{figure}{0}
        \setcounter{table}{0}
        \setcounter{lemma}{0}
        \setcounter{theorem}{0}
        \setcounter{corollary}{0}
        \setcounter{definition}{0}
        \setcounter{equation}{0}
        \renewcommand{\thefigure}{\Alph{appendixc}.\arabic{figure}}
        \renewcommand{\thetable}{\Alph{appendixc}.\arabic{table}}
        \renewcommand{\theappendixc}{\Alph{appendixc}}
        \renewcommand{\thelemma}{\Alph{appendixc}.\arabic{lemma}}
        \renewcommand{\thetheorem}{\Alph{appendixc}.\arabic{theorem}}
        \renewcommand{\thedefinition}{\Alph{appendixc}.\arabic{definition}}
        \renewcommand{\thecorollary}{\Alph{appendixc}.\arabic{corollary}}
        \renewcommand{\theequation}{\Alph{appendixc}.\arabic{equation}}
        \noindent{\tenbf Appendix \theappendixc #1}\par\vspace{5pt}}
\newcommand{\subappendix}[1] {\vspace{12pt}
        \refstepcounter{subappendixc}
        \noindent{\bf Appendix \thesubappendixc. {\kern1pt \bfit #1}}
	\par\vspace{5pt}}
\newcommand{\subsubappendix}[1] {\vspace{12pt}
        \refstepcounter{subsubappendixc}
        \noindent{\rm Appendix \thesubsubappendixc. {\kern1pt \tenit #1}}
	\par\vspace{5pt}}

\newcommand{\textlineskip}{\baselineskip=13pt}
\newcommand{\smalllineskip}{\baselineskip=10pt}

\def\eightcirc{
\begin{picture}(0,0)
\put(4.4,1.8){\circle{6.5}}
\end{picture}}
\def\eightcopyright{\eightcirc\kern2.7pt\hbox{\eightrm c}} 

\newcommand{\copyrightheading}[1]
	{\vspace*{-2.5cm}\smalllineskip{\flushleft
	{\footnotesize Modern Physics Letters A, #1}\\
	{\footnotesize $\eightcopyright$\, World Scientific Publishing
	 Company}\\
	 }}

\newcommand{\publisher}[2]{{\begin{center}\footnotesize\smalllineskip 
	Received #1\\
	Revised #2
	\end{center}
	}}

\def\abstracts#1#2#3{{
	\centering{\begin{minipage}{4.5in}\baselineskip=10pt\footnotesize
	\parindent=0pt #1\par 
	\parindent=15pt #2\par
	\parindent=15pt #3
	\end{minipage}}\par}} 

\newcommand{\bibit}{\nineit}

\renewenvironment{thebibliography}[1]
	{\frenchspacing
	 \ninerm\baselineskip=11pt
	 \begin{list}{\arabic{enumi}.}
        {\usecounter{enumi}\setlength{\parsep}{0pt}     
	 \setlength{\leftmargin 12.7pt}{\rightmargin 0pt} %FOR 1--9 ITEMS
         \setlength{\itemsep}{0pt} \settowidth
	{\labelwidth}{#1.}\sloppy}}{\end{list}}

\newcounter{itemlistc}
\newcounter{romanlistc}
\newcounter{alphlistc}
\newcounter{arabiclistc}

\newcommand{\fcaption}[1]{
        \refstepcounter{figure}
        \setbox\@tempboxa = \hbox{\footnotesize Fig.~\thefigure. #1}
        \ifdim \wd\@tempboxa > 5in
           {\begin{center}
        \parbox{5in}{\footnotesize\smalllineskip Fig.~\thefigure. #1}
            \end{center}}
        \else
             {\begin{center}
             {\footnotesize Fig.~\thefigure. #1}
              \end{center}}
        \fi}

\newcommand{\tcaption}[1]{
        \refstepcounter{table}
        \setbox\@tempboxa = \hbox{\footnotesize Table~\thetable. #1}
        \ifdim \wd\@tempboxa > 5in
           {\begin{center}
        \parbox{5in}{\footnotesize\smalllineskip Table~\thetable. #1}
            \end{center}}
        \else
             {\begin{center}
             {\footnotesize Table~\thetable. #1}
              \end{center}}
        \fi}

\def\@citex[#1]#2{\if@filesw\immediate\write\@auxout
	{\string\citation{#2}}\fi
\def\@citea{}\@cite{\@for\@citeb:=#2\do
	{\@citea\def\@citea{,}\@ifundefined
	{b@\@citeb}{{\bf ?}\@warning
	{Citation `\@citeb' on page \thepage \space undefined}}
	{\csname b@\@citeb\endcsname}}}{#1}}

\newif\if@cghi
\def\cite{\@cghitrue\@ifnextchar [{\@tempswatrue
	\@citex}{\@tempswafalse\@citex[]}}
\def\citelow{\@cghifalse\@ifnextchar [{\@tempswatrue
	\@citex}{\@tempswafalse\@citex[]}}
\def\@cite#1#2{{$\null^{#1}$\if@tempswa\typeout
	{IJCGA warning: optional citation argument 
	ignored: `#2'} \fi}}

\def\@refcitex[#1]#2{\if@filesw\immediate\write\@auxout
	{\string\citation{#2}}\fi
\def\@citea{}\@refcite{\@for\@citeb:=#2\do
	{\@citea\def\@citea{, }\@ifundefined
	{b@\@citeb}{{\bf ?}\@warning
	{Citation `\@citeb' on page \thepage \space undefined}}
	\hbox{\csname b@\@citeb\endcsname}}}{#1}}

\def\@refcite#1#2{{#1\if@tempswa\typeout
        {IJCGA warning: optional citation argument
	ignored: `#2'} \fi}}

\def\refcite{\@ifnextchar[{\@tempswatrue
	\@refcitex}{\@tempswafalse\@refcitex[]}}

\def\pmb#1{\setbox0=\hbox{#1}
	\kern-.025em\copy0\kern-\wd0
	\kern.05em\copy0\kern-\wd0
	\kern-.025em\raise.0433em\box0}

\def\fnm#1{$^{\mbox{\scriptsize #1}}$}
\def\fnt#1#2{\footnotetext{\kern-.3em
	{$^{\mbox{\scriptsize #1}}$}{#2}}}

\def\fpage#1{\begingroup
\voffset=.3in
\thispagestyle{empty}\begin{table}[b]\centerline{\footnotesize #1}
	\end{table}\endgroup}

\def\runninghead#1#2{\pagestyle{myheadings}
\markboth{{\protect\footnotesize\it{\quad #1}}\hfill}
{\hfill{\protect\footnotesize\it{#2\quad}}}}
\font\tenrm=cmr10
\font\tenit=cmti10 
\font\tenbf=cmbx10
\font\bfit=cmbxti10 at 10pt
\font\ninerm=cmr9
\font\nineit=cmti9

\font\eightrm=cmr8

\def\qed{\hbox{${\vcenter{\vbox{			%HOLLOW SQUARE
   \hrule height 0.4pt\hbox{\vrule width 0.4pt height 6pt
   \kern5pt\vrule width 0.4pt}\hrule height 0.4pt}}}$}}

\renewcommand{\thefootnote}{\fnsymbol{footnote}}	%USE SYMBOLIC FOOTNOTE

\begin{document}

\runninghead{Anisotropy in the Propagation of Radio Wave 
Polarizations $\ldots$} {
Anisotropy in the Propagation of Radio Wave 
Polarizations $\ldots$}

\normalsize\textlineskip
\thispagestyle{empty}
\setcounter{page}{1}

\copyrightheading{}			%{Vol. 0, No.0 (1992) 000--000}

\vspace*{0.88truein}

\fpage{1}
\centerline{\bf ANISOTROPY IN THE PROPAGATION OF RADIO POLARIZATIONS  }
\vspace*{0.035truein}
\centerline{\bf FROM COSMOLOGICALLY DISTANT GALAXIES} 
\vspace*{0.37truein}
\centerline{\footnotesize PANKAJ JAIN}
\vspace*{0.015truein}
\centerline{\footnotesize\it Physics Department, I.I.T. Kanpur}
\baselineskip=10pt
\centerline{\footnotesize\it  Kanpur, India - 208016}
\vspace*{10pt}
\centerline{\footnotesize JOHN P. RALSTON}
\vspace*{0.015truein}
\centerline{\footnotesize\it Department of Physics and Astronomy, Kansas University}
\baselineskip=10pt
\centerline{\footnotesize\it Lawrence, KS-66045, USA}
\vspace*{0.225truein}
\publisher{(received date)}{(revised date)}

\vspace*{0.21truein}
\abstracts{ Radiation traversing the observable universe provides powerful
ways to probe anisotropy of electromagnetic propagation. A controversial recent
study claimed a signal of dipole character.  Here we test a new and independent
data set of 361 points under the null proposal of {\it statistical
independence}
of linear polarization alignments relative to galaxy axes, versus angular
positions.  The null hypothesis is tested via maximum likelihood analysis of
best fits among numerous independent types of factored distributions.  We also
examine single-number correlations which are parameter free, invariant under
coordinate transformations, and distributed very robustly.  The statistics are
shown explicitly not to depend on the uneven distribution of sources on the
sky.
 We find that the null proposal is not supported at the level of less than 5\%
to less than 0.1\% by several independent statistics.  
The
signal of correlation violates parity, that is, symmetry under spatial
inversion, and requires a statistic which transforms properly. The data
indicate
an axis of correlation, on the basis of likelihood determined to be $[{\rm
R.A.}=(0^{\rm h},9^{\rm m}) \pm (1^{\rm h},0^{\rm m})$, ${\rm Decl.} = -1^o\pm
15^o]$. }{} {}

\vspace*{1pt}\textlineskip
\textheight=7.8truein
\setcounter{footnote}{0}
\renewcommand{\thefootnote}{\alph{footnote}}

\section{Introduction}
\noindent
The orientation of linear radio polarizations emitted by
cosmologically
distant galaxies has a consistent relation with the galaxy symmetry axis.
Exceedingly small physical effects accumulate during propagation, which
conventional measurements can directly probe.  Thus electromagnetic radiation
traversing the observable universe can detect subtle forms of cosmological
anisotropy. A signal with dipole character was claimed recently$^1$ 
 from an analysis of published radio data. Analysis found an
``anisotropy axis" $\vec s_{\rm NR}=(21^{\rm h}\pm 2^{\rm h}, 0^o\pm 20^o$)
governing orientation of polarization of the radio signals varying in a
coherent
way across the dome of the sky. The origin of this behavior is not clear, and
may or may not indicate a fundamental anisotropy on a scale larger than
previously found in cosmology.

There is a long history of puzzling observations. Beginning in the 1960's
observers noticed that Faraday-subtracted polarizations were distributed in
peculiar ways relative to the source axes. In 1982 Birch$^2$ empirically observed a
coherent angular anisotropy in the off-sets of the polarization and galaxy
axes,
using a data set of 137 points.  Birch's statistical methods were questioned,
but more sophisticated studies$^{3,4}$
 confirmed surprisingly strong signals in Birch's data.  The statistics
were not consistent with isotropy at 99.9\% and 99.98\% confidence levels,
respectively.  One of the same groups$^{4,5}$
 went on to create an independent set of 277 points and simultaneously 
introduced a different statistical measure. They obtained no signal in
this set and dismissed Birch's results. 
This left unresolved the puzzling fact
that his data had contained a signal at such a high level of statistical
significance. When Nodland and Ralston,$^1$ initially unaware of Birch's$^2$
work, independently found a statistically significant signal in an independent
set of 160 points, criticisms focused on proposing different statistical
baselines$^{6,7,8}$
and again claimed to find no signal of anisotropy.  The question of systematic
bias in such data had been raised by the authors$^1$
(henceforth $NR$) and earlier$^9$ regarding Birch.

Here we report analysis of a considerably larger data set which contains 361
points.  We have taken into account criticisms and experience from earlier
work,
and used the most robust statistical methods available. New progress has been
made by paying close attention to the symmetries of the problem. The usual
expectation of {\it independence} of the polarization and sky angular
coordinates, or ``uncorrelated isotropy'', happens to represent a definite
symmetry, which is that the distribution factors.  The classic scientific
method
becomes applicable: we can test isotropy as a clean hypothesis and see if
it can
be ruled out, which is immensely powerful.
 We use generic methods to represent the correlations,
emphasising the symmetry that they are {\it odd} in the polarization
variable at
hand, which is a consequence of parity (spatial inversion) symmetry.$^{10}$
 This simple point resolves many apparent discrepancies
between the
previous studies. Rather than being at odds with one another, all the facts are
now found to be consistent; we know of nothing in contradiction to our
conclusions.

The data collects variables from cosmologically distant galaxies, as
compiled in
the literature.$^{2,4,5,11,12,13}$
The data
set by
$NR$ reproduced that of Carroll et al$^{13}$ except for a half-dozen
corrections
from the original literature. The compilation of Eichendorf and Reinhardt,$^{11,12}$
available on the NASA-ADC archives, contain numerous sources for which the
position angle of the source is listed. We obtained the polarization angle for
these sources from Simard-Normandin et al$^{14}$  for all the sources for which
they were available. We compiled a total of 152 data points in this fashion.
Taking these as our primary data set we added any distinct data points
contained
in Bietenholz$^5$, making a total of 313 points.  Data points were regarded
as distinct if they had different Right Ascension, and differed in Declination
by more than one degree (which can be attributed to change in convention).
This
set was further combined with the $NR$ and remaining distinct points of the
Birch data, in that order, making a total of 361 data points. In combining
these
different data sets, we verified that the polarization off-set values for
points
with coincident Right Ascension and Declination did not differ by more than a
few degrees for most of the data. Specifically we found that the disagreement
exceeded $5^o$ only for very few points, which if deleted made no difference to
our final results.  
We also verified
consistency using a newer 1988 compilation by Broten et al.$^{15}$
 The only exception to this rule was found for Birch's data: here
the disagreement with other compilations was found to be larger, but still
tolerable.  All results we report are consistent, and no combination of any
large set gave results significantly different from any other.  
The absence of information available to us on Birch's $RM$ values,  
plus the possibility of
discrepancies in that data, led us to give results both with and without
Birch's
data.  In Figure 1 we show the angular distribution of data, which naturally is
not isotropic due to the zone of avoidance and dominance of Northern Hemisphere
measurements.  We will exhaustively show that the angular distribution is
not an
issue and cannot be confused with correlation.

\medskip

\begin{figure}[t,b]
\hbox{\hspace{0em}
\hbox{\psfig{file=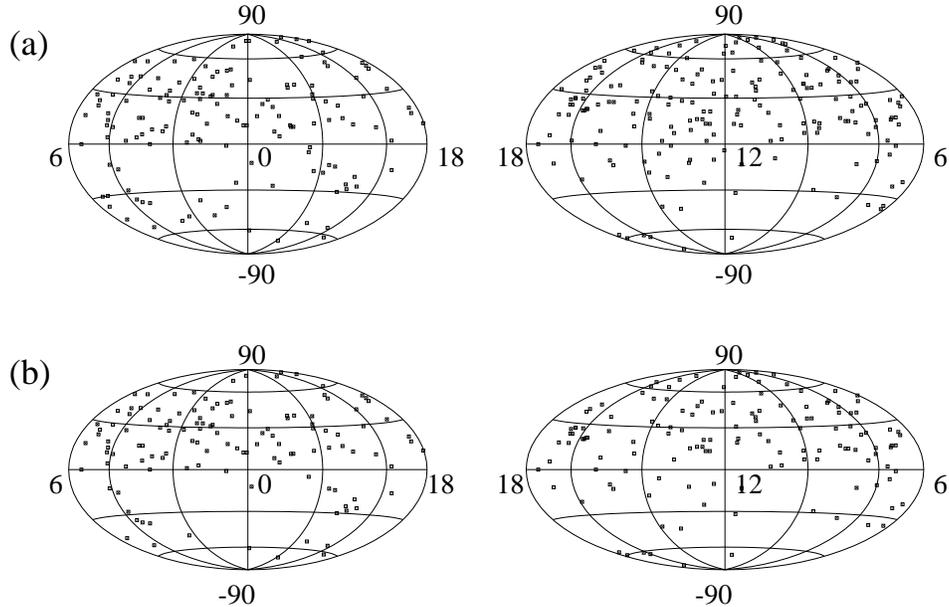,height=8cm}}}
\caption{
 Aithoff-Hammer equal-area plots of
the distribution of sources on the dome of the sky, in the equatorial
coordinates of the data used. The distribution is somewhat non-uniform due to
the zone of avoidance and dominance of Northern Hemisphere measurements.  (a)
The distribution of the full data set of 332 points, excluding the 29 extra
points contained in Birch's compilation. Adding Birch's data makes the set even
more uniform. (b) The distribution of the same data set after the cut on
rotation measure, $|RM-\overline{RM}| >6$.  Any non-uniformity of the angular
distribution is taken into account in all statistics reported.
}
\end{figure}

\medskip

The observables listed for galaxy $i$ include a major axis orientation angle
$\psi_i$, a linear polarization angle $\chi_i$, and the angular coordinates of
the galaxies on the sky. Other variables may include a resolution parameter,
degree of polarization, and the Faraday rotation measure $RM$. The rotation
measure is the slope of plots of measured polarization angle versus
wavelength-squared.  This is known to measure intervening magnetized plasma
parameters. A-priori, $RM$ has nothing to do with the variable $\chi$, which is
the polarization angle after Faraday rotation is subtracted.  However we have
retained this variable, which seems to be informative. Consistent with
restricting the study to uncorrelated isotropy, we integrate over the redshift,
which happens to be incomplete in the data set in any event.  We let
$\beta=\chi-\psi$ be the angle between the plane of polarization and the
symmetry axis of the source.  The variables $ \chi$ and $\psi$ are
determined up
to a multiple of $\pi$; $\beta$ runs from $-\pi$ to $\pi$.$^{10}$
 To deal with the $\pi$ ambiguity of polarization and axis
measurements,
one can map $\beta \rightarrow Y(\Omega)$, where $\Omega$ is a variable defined
on twice the interval.  A popular map is ``{\it Map 1}'', $\Omega_1(\beta) = 2
\beta$.  The function $Y$ is represented by a Fourier series with periodicity
$2\pi$, assuring that the transformation $\beta \rightarrow \beta^\prime =
\beta
\pm \pi$ leaves $Y(\Omega)$ invariant.  The first Fourier components create a
2-component vector-like object $\vec Y(\Omega) = (\cos(\Omega), \sin(\Omega))$.
When the components of $\vec Y(\Omega)$ are used in statistical analysis, there
is naturally a Jacobian factor which represents the choice of {\it Map}.  By no
means, then, is {\it Map 1} sacred, and other maps are discussed below. The
angular positions on the dome of the sky are mapped into their
3-dimensional
Cartesian vector positions $\vec X$ on a unit sphere.  Since we do not model
this distribution, but take it from the data, this standard map is adequate.
When coordinate origins are changed, the components of $\vec X$ transform by
standard rules; one can go on to make nicely transforming distributions and
tensor correlations. The two choices measuring $\chi$ relative to $\psi$ or
$\psi+\pi/2$ correspond to $\vec Y\rightarrow -\vec Y$.  This does not mix
the 2
components of $\vec Y$, which will be called ``even'' (for $\cos(\Omega)$) and
``odd'' (for $\sin(\Omega)$) following the transformation property of being
even or odd, respectively, under parity (spatial inversion). As discussed in 
detail elsewhere,$^{10}$ functions of the offset angles have the corresponding 
parity if they are even or odd functions of $\beta$, as intuitively evident 
from the handed ``sense of twist'' a parity-odd quantity conveys.   
The invariant correlations discussed below avoid
any question of coordinate origin (either in polarization quantities or in
angular positions on the dome of the sky) by being totally independent of the
choice of angular origin.

The standard assumption of statistical independence corresponds to a
distribution $g(\Omega,\vec X)=h(\Omega) f(\vec X )$.  This is a very broad
class of distributions, with $h(\Omega)$ and $f(\vec X )$ completely
unrestricted, which nevertheless has symmetries allowing it to be tested. All
statistics will be compared to baselines using the actual distribution of the
data $f(\vec X)$ on the dome of the sky in Monte Carlo simulations.  Statistics
based on assuming independence of polarizations and positions will be compared
with a simple correlated ansatz of the form $h(\Omega) C(\Omega, \vec X) f(\vec
X)$. The case $C=1$ reduces to the uncorrelated case.

\section{Methods}
\noindent
We report both raw statistics and also ``P-values'', defined as the integrated
probability for the null proposal to give an equal or larger statistic.  We
also
use ``confidence levels'' defined to be 1-P.  We will present 2 methods testing
for independence:

\subsection{Invariant Correlations} 
\noindent 
``Summary'' correlations are
single number quantities which do not necessarily probe all types of
relationship.  We use these for preliminary tests, and to support the more
sensitive likelihood analysis with complementary information.

Following Jupp and Mardia$^{16}$ ($JM$), multivariable $p\times q$ matrices are
defined via $m_{ij}^{XY}=\sum\limits_\ell (X- \bar X)^\ell_i (Y- \bar
Y)^\ell_j\; $, with $(m^ {XX})^{-1}_{ij}$etc. similarly defined.  Note that all
vectors used in the correlation matrices have their means in the sample,
denoted
by ``bar", subtracted.  The invariant JM correlation test statistics for $n$
data points are $n\rho^2_{p\times q}$ where $\rho_{p\times q}^2 =Tr[(m^{XX})^{-
1}\cdot (m^{XY})\cdot (m^{YY})^{- 1}\cdot m^{YX}]$ where Tr denoted the
trace of
the matrix. An important and simple feature of $\rho^2_{p\times q}$ is that
$\rho^2_{p\times q}=0$ when the distribution (of any kind) is uncorrelated.
The
$JM$ correlations, in addition, satisfy $0\leq \rho^2\leq min (p, q)$,
achieving
maximum for perfect correlation, and distribution of $\rho^2_{p\times q}$
independent of the marginal distributions. Note that $\rho^2_{p\times q}$ is
invariant under separate rotations of origins of galaxy axes and polarizations,
or orthogonal transformations of sky coordinates, and also does not involve any
parameters.  We have $p=3, q=1$ for the correlation of sky positions $\vec X$
and the separate even-or odd-parity$^{10}$ $\beta$
representations $\vec Y_1, \vec Y_2$, respectively.  The distribution of
$n\rho^2_{3\times 1}$ is known to be $\chi^2_3$ for large $n$.  We verified
this
with extensive Monte Carlo simulations, confirming that the statistics are
quite
robust.  In fact, the statistics of $n\rho^2$ were devoid of detectable
dependence on different marginal distributions, which included the flat
distribution, the von Mises, and shuffled distributions described below. A
graph
of Monte Carlo generated distributions from 10, 000 trials is shown in
Figure 2,
showing excellent agreement with the $\chi^2_3$ distribution, while consistency
of P-values was also checked to eliminate the possibility of long tails.

\medskip

\begin{figure}[t,b]
\hbox{\hspace{0em}
\hbox{\psfig{file=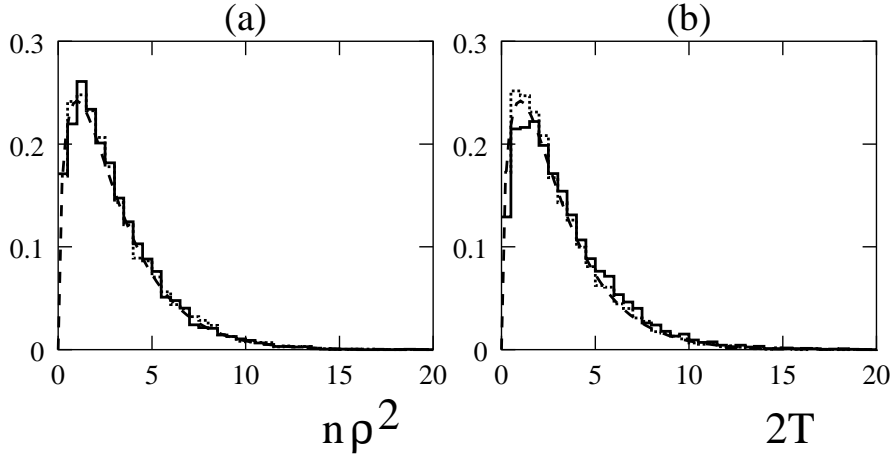,height=6cm}}}
\caption{
  (a) Distributions of the $ 1\times 3 $ JM statistic discussed in the
text.  Dashed curve: the distribution $\chi^2_3$ predicted by JM. Solid 
histogram: the distribution generated by shuffling the full data set's $\beta$
and $\vec X$ coordinates to create an uncorellated null distribution with the
same marginal $\beta$ and $\vec X$ distributions. Dotted histogram: same as the
dashed case, but with the data set cut to $ | RM - \overline{RM}| >6$.  The
Monte Carlo distributions are indistinguishable from the analytic prediction,
and demonstrate that the statistic does not depend on the distribution of
points
on the sky. (b) Distribution of the statistic $2T$, where $T$ is the difference
of likelihoods of the correlated model and the null model. The dashed curve is
the distributions $chi^2_3$ predicted on the basis of asymptotic theory.  The
solid histogram is the result obtained by Monte Carlo simulation for the case
of the full data set of 332 points. The dotted histogram is the result of the
simulation obtained after the cut $|RM-\overline{RM}|>6$. The $\beta$ values
were generated by shuffling the $\beta$ values of the original measured data
sets. For the case of the cut $|RM-\overline{RM}|$, it was found that in no
cases (either for the Jupp Mardia analysis and the Likelihood analysis) did
P-values generated by the Monte Carlo using 10,000 random sets exceed P-values
reported, a consistency check eliminating any long tails.
}
\end{figure}

\medskip

The $JM$ correlations are simple canonical summaries inspired by Gaussian
statistics.  For angular variables they are limited, but suited to sense
particular linear relations between $\vec X$ and $\vec Y$, namely orthogonal
rotations times projections.$^{10}$ Other natural coherent
correlations can unfortunately yield zero -- for example, the $JM$ correlation
fails to see that vector $\hat\theta$ in spherical polar coordinates is
correlated with a uniform distribution of angular position. Such
limitations are
expected because correlation might take myriad forms orthogonal to a summary
statistic's power to probe. A small $JM$ test correlation does not prove
isotropy for this reason. Nevertheless a sufficiently large $JM$
correlation can
logically rule out isotropy with a definite statistical significance.

\subsection{Independence of Statistics on the Sky Distribution}
\noindent 
Not
only are the JM statistics independent of the distribution of $\beta$, but for
large $N_{data}$ also independent of the distribution of data points on the
dome
of the sky.  This is because they are tests for independence, not tests of
particular distributions.  The formal proof is given by JM.$^{16}$ 
We also tested the distribution directly with the data and with Monte
Carlo trials.  Specifically (Figure 2), we shuffled the full data sets $\beta$
and $\vec X$ values to create an uncorrelated null distribution having the same
marginal distributions in $\beta$ and $\vec X$ as the data.   Shuffling is
preferred here, eliminating any question of possible imperfections of models of
the uneven sky distribution, which were never used anywhere in our analysis.
Runs with 10, 000 copies of shuffled data gave excellent agreement with the
analytic distribution known to be $\chi^2_3$.  Next we repeated the procedure
with the large $RM$ cut, to be discussed.  The distribution was again the same
(Figure 2) and in excellent agreement. This is another demonstration that there
is no issue of the non-uniformity of the sky distribution affecting the
statistics.  There have been many misunderstandings of this basic point. We
went
on to try an experiment with the same number points, and using the 
$\beta$ values from the data, but with $\vec X$ values randomly distributed
in extremely restricted angular regions, within a cluster of 
of half-angle
$\pm$
7.5 degrees extent in right ascension and declination (Figure 3). We also
examined the statistic's 
distribution with $\vec X$ values restricted to 
two oppositely oriented clusters of the same
size,
and restricted to 
an equatorial belt of the same angular width. In each case 10, 000 Monte
Carlo runs were made to generate the distributions. The distribution of each
case was indistinguishable from the others, and the same as the predicted 
$\chi^2_3$.

\begin{figure}[t,b]
\hbox{\hspace{2em}
\hbox{\psfig{file=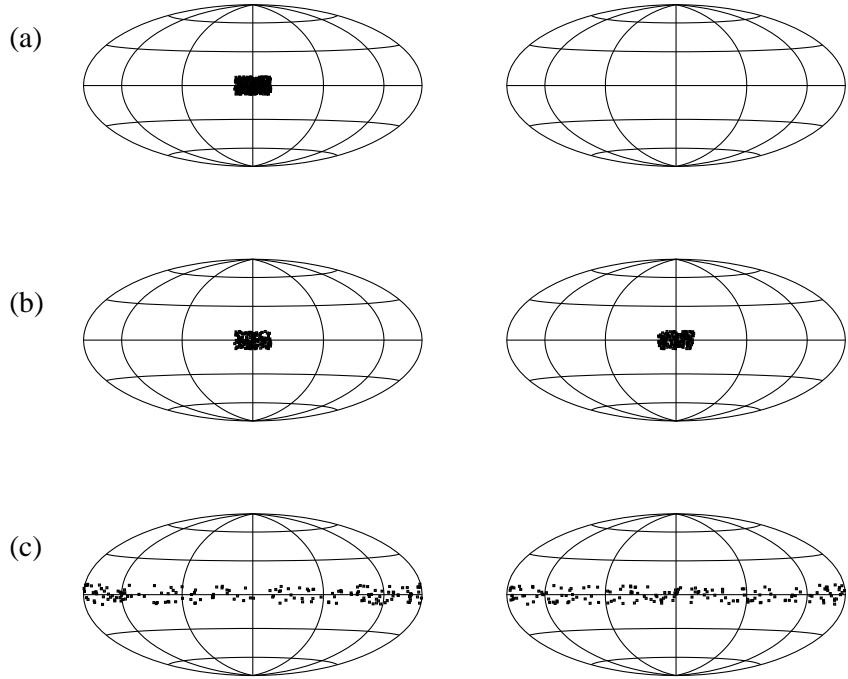,height=9cm}}}
\caption{
 Aithoff-Hammer plots of Monte Carlo simulation with randomly
generated
sources restricted to small angular patches on the dome of the sky.  (a) A
single patch with half-width in latitude and longitude of 7.5  degrees. (b) Two
patches on opposite sides of the sphere with the same angular width as (a).
(c)
A belt of the same angular width centered on the equator. The JM$1\times 3$
statistic for each case was evaluated for 10,000 randomly generated trials.
Distributions of the test statistic were in excellent agreement with the
predicted distribution $\chi^2_3$ in each of these three cases, and
indistinguishable from those of Figure 2.   The angular distributions,
being far
less isotropic than that of the data (compare Figure 1), serve to verify
independence of the statistic on the sky distribution.
}
\end{figure}

\medskip

\subsection{Likelihood Analysis} 
\noindent 
The more sensitive method is
likelihood analysis, a straightforward contest between model distributions. The
likelihood $L$ is defined to be the product of the normalized distribution
$f(z)$ evaluated at the points $z_i$ of the data: $L  = \prod_i f(z_i)$.

The logarithm of the likelihood is maximized as a function of distribution
parameters.  Twice the difference of maximized log-likelihoods ($2T$) between a
factorized null model and a model containing correlation is distributed like
$\chi^2_m$, where $m$ is the difference in the number of parameters of the
models.  The effects of parameters on statistical significance is thereby taken
into account.   As mentioned earlier, each study we present takes into account
the distribution of the sample on the dome of the sky explicitly by taking
$f(\vec X )$ directly from the data.  We checked the results by extensive Monte
Carlo statistical simulations (described below). To create the most optimistic
definition, we conservatively allowed the null hypothesis to be whatever is the
best {\it a posteriori} champion for the null hypothesis.  This is a
conservative bias. Nine independent functional forms with many free parameters
were explored to represent $h(\Omega)$. As for the correlated distribution
$C(\Omega, \vec X)$, we did not make extensive searches, but used a
standard map
and the map produced by the best null fit in the correlation ansatz of Kendall
and Young,$^3$ both of 
which are odd functions under $\beta\rightarrow -\beta$ representing 
odd-parity.

\section{Invariant Correlations}
\noindent
Following $NR$ (a signal odd in $\beta$), for our first preliminary study we
apply the odd-parity {\it Map 1} $JM$ correlations to the full data set, and to
cuts based on rotation measure ($RM$).  An odd parity signal was also reported
earlier by Birch$^2$ for a smaller 137 point data set and verified later by
Kendall and Young.$^3$  The motivation for cutting on $RM$ is that this is
the only variable we have to give a symmetric and unbiased cut.   A further,
physics-based motivation$^{17}$ appears in the {\it
Discussion}.

Results for the {\it full data set}, 361 points, with {\it Map 1} are reported
in {\it Table 1}.  In particular we find $n \rho^2_{1\times 3}=11.15$.  This
indicates a statistically significant indication of correlation with this
measure, with a P-value of 1\%.  The usual isotropic uncorrelated hypothesis is
ruled out at 99\% confidence level.  If we exclude the extra points from 
Birch's sample, we are left with a 332 point data set for which
$n\rho^2_{1\times 3} = 8.17$ ($P-value = 4.3\%$), which also shows evidence for
correlation.

\begin{table}[htbp]
\tcaption{Correlation test statistics $n\rho^2$ as defined by Jupp and
Mardia, a
scale and rotationally invariant trace of products of $1 \times 3$ correlation
matrices, along with maximum log-likelihood test statistics $2T$. P--values are
the probability of fluctuations in the null distribution to equal or exceed the
data's statistic. Results are given for two different choices of mappings
$\beta
\rightarrow \vec Y(\Omega)$ discussed in text, with {\it Map 1} and {\it Map 3}
corresponding to  $\Omega_1 = 2\beta$ and $\Omega_3= 2\beta + \nu \sin(2\beta)$
respectively.  Best fits to the null hypothesis selected {\it Map 3} over {\it
Map 1}. P-values indicated by $*$ use $\chi^2_5$.}
\centerline{\footnotesize\smalllineskip
\begin{tabular}{cccccc}\\ \hline {\bf
Data}
&{\bf no. of} & {\bf Map 1} & {\bf Map 1} & {\bf Map 3} ($\nu=-1$)&{\bf Map
3}\\
{\bf set} &{\bf points} &{\bf  $n\rho^2$} (P) &2T (P) & {\bf  $n\rho^2$} (P) &
2T (P)\\ \hline && & & & \\ Full & 361 & 11.15 ($1.1$\%) & 8.40 (3.8\%)  &
12.77
($0.5$\%) & 11.22 (1.1\%) \\ && & & &  \\ Excluding & &&&&\\ Birch & 332 & 8.17
($4.3\%$) & 5.54 (14\%) & 10.10 (1.8\%) & 8.6 (3.5\%) \\ && & & &  \\
RM$<0$ & &
 &&&\\ Excluding & 153 & 14.34 (0.2\%) & 11.90 (0.8\%) & 17.25 (0.06\%) & 17.26
(0.06\%) \\ Birch &&&&& \\ && &&&\\ RM$\ge 0$ & &&&& \\ Excluding & 179 & 5.37
(15\%) & 4.62 (20\%) & 3.60 (30\%) & 0.44 (93\%) \\ Birch &&&&& \\ && & & & \\
$|RM-\overline{RM}|>6$ & &&&& \\ Excluding & 265 & 16.62 (0.5\%$^*$) & 15.56
(0.8\%$^*$) & 22.66 (0.04\%$^*$) & 21.62 (0.06\%$^*$) \\ Birch &&&&& \\ && & & & \\
\hline
\end{tabular}} \end{table}

We now turn to cuts based on rotation measure ($RM$), which we
apply to
the 332 point data set (excluding Birch's compilation since we did not have
the RM for this set). We examined the regions of $RM< 0$ (153 points) and
$RM\ge
0$ (179 points) separately, which in an unbiased set would be a search
between 2
alternatives.  Sticking to the strict cuts, the region of $RM \ge 0$ does not
show evidence of correlation: $n\rho^2_{1\times 3}=5.37$. However we find that
the region of $RM<0$ shows a highly significant {\it Map 1} correlation:
$n\rho^2_{3\times 1, odd} = 14.3$  (P-Value = $3\times 10^{-3}$).  For those
accustomed to Gaussian 2-sided statistics, this corresponds to a deviation at
greater than the $3\sigma$ level, or a 99.7\% confidence level that the null
does not fit the data.

\medskip The question may arise whether the point $RM=0$ is unique. Varying the
cut location, a significant correlation is quite persistent for cuts retaining
$RM< 0$.  By varying the cut in the region $RM< -20$ to $RM< 7 $, we find
statistically significant ( 95\% or better) rejection of the null
hypothesis for
each of the 27 ways to choose the cut. The number of points varies quite
rapidly
in this region.  Meanwhile the statistic nicely follow proportionality to the
number of points $n_{cut}$, that is $n_{cut}\rho^2_{cut} \approx
(n_{cut}/n_{RM<
0}) \times n_{RM< 0}\rho^2_{RM< 0}$.  This is non-trivial and not consistent
with fluctuations ($\rho^2$ is a nonlinear function) but rather consistent with
finite $\rho^2$ being an intrinsic property of the sample.  From this
observation one projects that statistical significance should be lost for cuts
on $RM < -20$, due to decreasing the number of points.   Cutting the positive
side of the origin of $RM$, P-values exceed  5 \% for cut values of $RM_{\rm
min} > 7$.

The significant correlation for the set $RM < 0$ might be misinterpreted.  The
question arises whether this might be due to a decorrelated strip in the
complementary region $RM \ge 0$ (as opposed to something special about $RM <
0$). Indeed the $RM$ distribution is not Gaussian but instead has a shifted
mean
$\overline{RM}= 6$, with a big spike at  the mean plus or minus $6-10$ units.
The central spike region, occuring entirely outside the $RM < 0$ region,
appears
to be uncorrelated. The balance of the data is then highly correlated {\it for
both positive and negative $RM$}.  This is quite a striking phenomenon. This
effect is most clear in the likelihood analysis (below): but it is also visible
with $n\rho^2_{1\times 3, odd}=16.6 $, $P-value < 8\times 10^{-4}$
for the
region excluding $\overline{RM} \pm 6$. 
In order to account for the 2 parameters needed in specifying the cut,
namely the mean and the width of the excluded region, we conservatively use
$\chi^2_5$ distribution to evaluate the probability, 
and find a P-value of $5\times 10^{-3}$.
Again this is not very sensitive to the
method of cut: a statistically significant rejection of the null hypothesis at
99\% confidence level occurs for every one of the 13 ways to exclude
$\overline{RM} \pm 5$ to $\overline{RM}\pm 18$.

\section{Likelihood Analysis and the Map}
\noindent
A more specific method of testing isotropy is likelihood analysis. A typical
null distribution for angular variables is the von Mises ({\it vM}) form,
of the
type $$h(\Omega) f(\vec X) = constant\times\exp(k\cos(\Omega))f(\vec
X)\eqno (1)
$$ The distribution has its maximum at $\Omega=0$ or $\Omega=\pi$ for $k>0$ or
$k<0$, respectively. The location of the maximum can be translated by adding a
parameter $\Omega \rightarrow \Omega - \delta$.  As a generic ansatz odd in
$\beta$, we used the Kendall and Young$^3$ {\it (KY)} ansatz for
correlation:
it is $$C(\beta,\vec X) = \exp\lbrace \mu\vec X\cdot \vec s\sin(\Omega
)\rbrace\eqno (2)$$ The exponential forms are inspired by statistics, rather
than any deep physical considerations: they are exponentials of linear and
bilinear functions of $\vec X$ and $\vec Y$.  The correlation ansatz depends on
3 parameters: a measure of correlation $\mu$, constructed so that $C=1$ when
$\mu=0$, and 2 parameters locating a normalized $\vec s$. Lines of constant
probability are a linear relation between components of $\vec Y$ and
$cos(\gamma)$, where $\gamma$ is the polar angle between a source and $\vec s$.
To compare correlation\fnm{a}\fnt{a}{The shifted $vM$(Loredo et al$^8$), 
$g_{\rm
shifted-vM}(\Omega) =constant\times\exp(k\cos(\Omega- \theta(\vec X))$, where
$\theta(\vec X)$ is some correlating function, is an inappropriate test for
correlation. First, the model correlation is even-parity, 
and cannot fit an odd-parity
correlation we study here. Second, the marginal $\Omega$ distribution becomes
completely tied to the $\theta(\vec X)$ distribution: a uniform
$\theta(\vec X)$
distribution, e.g., generates a {\it flat} $\Omega$ marginal distribution,
incompatible with the data.}  we separately maximized the log-likelihoods $L_2$
(correlated fit) and $L_1$ (null fit), extracting the test statistic $2T
=2(L_2-L_2)$, which is distributed like $\chi^2_3$.

Applying the maximum likelihood analysis to the full data set of 361
points, one finds ({\it Map 1}) that $2T=8.40$, with a P-value of 3.8\% (or
96\%
confidence level evidence of dependence). In this case the best fit parameters
are ($k= -0.65$) (null), ($k=-0.66, \mu=0.39 $) (correlated), with axis
parameters $\vec s = \big[(1^{\rm h},16^{\rm m})\pm (2^{\rm h},0^{\rm m})\ ,\
40^o\pm 20^o\big].$

Having found that correlation fits significantly better than a standard
null, we
further examined several different {\it Maps} attempting to improve the null
fit. We investigated von Mises distributions depending on $\Omega_2= 2\beta -
\delta$, where $\delta$ is an arbitrary parameter. Other maps are:

\begin{tabular}{ll} Map 3:& $\Omega_3 = 2\beta + \nu \sin(2 \beta)$ \cr Map
4: &
$\Omega_4 = 2\beta+\nu\sin(2\beta) + \delta$\cr Map 5:& $\Omega_5 =
2\beta+\nu\sin(2\beta + \delta)$\cr Map 6:& $\Omega_6 =
2\beta+\nu\cos(2\beta)$\cr Map 7:& $\Omega_7 = 2\beta+\nu\cos(2\beta) +
\delta$\cr \end{tabular}

This gives 7 linearly independent maps in all.  We also used $\Omega_1$ in the
bimodal von Mises combination $\delta c_1 \exp[k_1\cos(2\beta)] +
(1-\delta) c_2
\exp[k_2 \cos(2\beta - \pi)]$ which has 3 parameters $\delta, k_1, k_2$, and in
the 2- parameter cardiod distribution$^{18}$  ${1\over \pi} (1 + k
\cos(\Omega_1-\delta))$. We have then nine independent functional ways
trying to
make the null proposal fit the data.

As a result we were able to improve the likelihood of the null fit
significantly, with {\it Map 3} producing by far the best fit, about 2.4
($2T=4.8$) units of likelihood higher than the $vM$ distribution while using
only one more parameter. This map (standard in ``circular statistics''
biological studies$^{19}$ on the 
swimming of fishes, escape of salamanders etc.)
 makes a more flattop or sharply peaked distribution
than the
$vM$, as definitely required here. To check for correlation we compare the same
map in the $KY$ correlation (Eq. 2).

Remarkably the improved null fits generate an even greater signal of
correlation.\footnote{In order to give the null the advantages, a common
conservative bias, we actually optimized all combinations of $\chi^2$
degrees of
freedom and likelihood values of the null distribution as compared to the
correlated case. For example, one can pit the single-parameter $vM$ with lower
likelihood against the more parameter {\it Map 3} hoping to improve the null's
significance using $\chi^2_4$. It fails. For another example, we conservatively
rejected a slightly better fit for the bimodal $vM$. By having more parameters
it presents a weaker case for the null proposal in likelihood comparisons.} The
results are summarized in {\it Table 1}; for completeness, parameters are
listed
in the text, while uncertainties are listed as cited.  The full data set, 361
points, shows $2L_2-2L_1= 2T=11.22$, a $P$-value of 1\%, or more than
$2\sigma$'s deviation from the expectations of an uncorrelated distribution.
This is another statistically significant indication of {\it dependence}.
Objectively, the uncorrelated isotropic assumption is sufficiently worse in
fitting the data that it is significantly disfavored. The best fit
parameters in
this case are ($k=-0.63, \nu=-0.50$) (null), ($k=-0.61, \nu=-0.67, \mu=0.47 $)
(correlated), with axis parameters, $\vec s = \left[(0^{\rm h},33^{\rm m})\pm
(1^{\rm h},40^{\rm m})\ , \ 34^o\pm 20^o\right]$. For the 332 point set, which
excludes Birch's compilation, we find $2T = 8.6$, ($P-value = 3.5\%$), with
best
fit parameters ($k=-0.62, \nu=-0.58$) null, ($k=-0.60, \nu=-0.75, \mu = 0.44$)
(correlated), with axis parameters, $\vec s = \left[(23^{\rm h},22^{\rm m})\pm
(1^{\rm h},40^{\rm m})\ , \ 29^o\pm 20^o\right]$. On the full data set  the cut
$RM< 0$ yields very significant evidence disfavoring the null distribution by
more than $3.5\sigma$, a P- value of $ 6 \times 10^{-4}$. If we exclude the
region $|RM -\overline{RM}| \le 6$ we find a spectacularly large correlation
with $2T = 21.62$ and parameters ($k=-0.60,
\nu=-0.37$) (null), ($k=-0.55, \nu=-0.86, \mu=0.79$) (correlated) with axis
parameters, $$\vec s = \left[(0^{\rm h},9^{\rm m})\pm (1^{\rm h},0^{\rm
m})\ , \
-1^o\pm 15^o\right]\ \ .$$

While the measures of significance are standard, we also verified the
likelihood
difference P-values by Monte Carlo calculations comparing the $2T$ value
for the
data where $\beta$ was generated randomly, in one method by shuffling and in
another method using our best fit distribution to the measured data.  The Monte
Carlo distributions are in excellent agreement and shown in Figure 2.

\section{Discussion}
\subsection{Parameters} 
\noindent 
The statistics of $2T$ take parameters into account.
The $P$-values represent probabilities for fluctuations from the null
distribution to appear correlated when fitting with any values of the
parameters
whatsoever. The parameter $\mu$ represents a strength of correlation in the
model ansatz, which is found to be relatively ``of order unity" compared to the
other parameters. The parameter $\vec s$ represents orientation of a normalized
axis with 2 degrees of freedom on the dome of the sky. The likelihood analysis
yields $\vec s$ parameters that tend to agree reasonably within errors with the
axis $\vec s_{NR}$ extracted in $NR$: $\vec s_{NR}=(21\pm 2$hrs, $0^o\pm
20^o$).
Variation is expected, of course, given the substantial differences in
statistical approach, and different data sets. The 2-quadrant procedure of
Nodland and Ralston$^1$ is quite distinct from both the likelihood
analysis and invariant correlations, making the near coincidence of axis surely
significant.  Some proposals for the axis orientation have been made earlier$^{20,21}$
and numerous theoretical
models,$^{13,22,23,24,25,26,27,28,29,30}$
interpretations$^{31,32,33,34,35}$
or related issues$^{17}$ have come up.

\subsection{Cuts} 
\noindent 
As emphasized earlier, the entire data set is
correlated. Yet the cuts on $RM$ produce such a significant result that they
cannot be dismissed. It is known that $RM$ is correlated with position on the
sky; thus $RM$ cuts may tend to select certain regions preferentially. Figure 1
shows visually only a small change in the angular distribution, however. Our
procedures take into account the population of sources on the dome of the sky,
so the results are not due to changes in this population. We verified this
extensively, using $\beta$ values from the data shuffled and assigned randomly
to Monte Carlo generated points on the dome of the sky. The angular regions of
the Monte Carlo data could even be restricted to patches far less isotropic
than
the data without detectable change in our statistic's distribution, 
as discussed earlier in the section on {\it Independence of Statistics on 
Sky Distribution}.
 We
reiterate, then, that the statistic does not depend on the angular distribution
of the data; unevenness of the sky distribution does not cause a correlation.
But for the question what is causing the signal, we cannot rule out the
possibility that different sky regions might be better correlated than others.
Resolving such questions would go beyond the scope of this paper.

\subsection{Other Methods} 
\noindent 
If $\beta$ were an angle depending
on the
angular coordinate system, transformed under change of angular origin
 by mixing the components of $\vec Y$,
then another statistic that might be examined is the mixed even-odd
parity combination
$n\rho^2_{2\times 3}$. As emphasized in Ref. [16] this quantity is free 
from any dependence on angular origin. While useful for some purposes, the
statistic is not appropriate for our study, because $\beta$ is already
invariant under change of origin. Equally important, representations of
different parity are mixed together in a non-linear way in 
$\rho^2_{2\times 3}$. As a result of combining many degrees of freedom,
the mixed parity $2\times 3$ statistic is distributed with 6 degrees
of freedom ($\chi^2_6$) compared to 3 degrees of 
freedom for the pure-parity $1\times 3$ correlations. It follows that
the statistical significance of a pure parity-odd (or pure parity-even)
correlation will not be properly evaluated, but greatly underestimated,
using $\rho^2_{2\times 3}$. This has been a point of confusion in the
literature, 
as Bietenholz and Kronberg$^{4,5}$ ($BK$) dismissed Birch's
claims on the basis of the mixed-parity $\rho^2_{2\times 3}$ statistic,
while simultaneously citing that Birch had a pseudovector effect. We
also were misled at first, and in preliminary work we confirmed 
$BK$'s calculations.
Surprisingly, the negative $RM$ region and center-$RM$ deleted
regions are so highly correlated that even on the basis of 
the mixed $n
\rho^2_{2\times 3}$ there is a strong signal: for the 131 points of $BK$'s set
with $RM< 0$, $n \rho^2_{2\times 3, Map 1} =19.48$, a confidence level of
99.7\%
that the set is correlated.  Consistent with all the other results, this data
set definitely has highly significant odd correlations $n\rho^2_{3\times 1,
odd}
=  11.50$  , P-Value = $9\times 10^{-3}$. There may also be mixed or 
even correlations:
$n\rho^2_{3\times 1, even} = 8.64$, (P-value=3.5\%); the existence of both
explains the large $n\rho^2_{2\times 3}$.  Otherwise we did not find
significant
{\it even} $JM_{1\times 3}$ correlations.

Another approach is to calculate the $1\times 1$ correlations of polarization
representations and sky position vectors projected into a subspace defined by a
fixed axis.  A similar (mixed parity) procedure was reported by Bietenholz,$^5$
but the method unfortunately involves searching for an axis parameter.
We studied this in the early stages, revealing large correlations when axes
were
fit. We have choosen not to make claims about that method here, due to the
difficulty of accounting for the effects of a parameter search, but as a
consistency check the result is meaningful.

More recent suggestions$^{6,8}$ to check sensitivity to the statistical baselines used
in $NR$  have already been
incorporated in more sophisticated ways by our analysis.  In another study
Carroll and Field$^7$  suggested a minimum chi-squared procedure, using the
Euclidean angular distances from the mean. The concept of an ``average angle"
defined by the arithmetic mean is then introduced.  As discussed by Batschelet$^{19}$
(see also Mardia$^{36}$, Fisher$^{18}$) the ``average angle" so defined has
caused problems in many fields.  It is a quantity which does not transform
properly, but instead depends on the convention used in binning angles.  The
same goes for the {\it rms} deviation from the mean, and we cannot use such
methods to obtain estimators of anything physical or statistical.

As a side note, we remark that even--odd symmetry is also useful for discussing
the null distribution.  More demanding than simple independence is
requiring the
marginal distribution to be an {\it even} function of $\beta$.  We did not
force
this, because a variety of effects, from bias to sampling statistics, might
cause the data to violate this symmetry, and the best fit null of any kind was
our goal. It seems very significant, then, that an unbiased marginal $\beta$
distribution fit delivers the {\it even} sharply -peaked ansatz, which is
another indication that the data is acting physically.

\subsection{Are There Loopholes?} 
\noindent 
Granted that the data is
surprising, a conservative interpretation is certainly prudent. A feature of
likelihood analysis, vigorously addressed earlier, is that there might exist
some unknown and better-fitting null distribution we were unable to find. To
close any loopholes one might ask for a further statistic not depending on
distribution models. Now the likelihood analysis obtained a sharply peaked {\it
Map 3}, with the parameter $\nu=-1$ indicated. Adopting this {\it Map} we
recalculated the odd--$1\times 3$ correlations.  Strongly consistent is that
with {\it Map 3} all correlations increase again.  Specifically ({\it Table
1}),
for the full data set (361 points) $n\rho^2_{3\times 1} =12.77$; for the data
excluding Birch's set (332 points) $n\rho^2_{3\times 1} = 10.10$;  for the cut
excluding the region $|RM-\bar{RM}| \le 6$, then $n\rho^2_{3\times 1}$ = 22.60
with the corresponding P-values of $ 5 \times 10^{-3}$, $2\times 10^{-2}$ and
$3\times 10^{-5}$ respectively. These particular P-values should be interpreted
with care because prior information was used. Yet the more sensitive {\it
Map 3}
correlations are useful because they are independent of any distribution model.
At better than 99.95 \% confidence level, the results show that no null
distribution exists to fit the data.

\subsection{Bias} 
\noindent 
Let us address the possibility of some
consistently imposed bias. We make no pretense to special qualifications but
have done what we can.  Regarding polarizations, besides using the data sets
already mentioned, we also checked consistency using a newer 1988 compilation.$^{15}$
These polarization measurements are largely
independent, scattering by a few degrees when sources coincide, as expected.
Redoing the entire analysis with these polarizations as ``primary'' for
coincident sources, the statistics are hardly changed, and the null is again
ruled out at the 99\% to 99.9\% confidence level.   Regarding the galaxy axes,
we also have an independent set from the work of Bietenholz.$^5$  Indeed
this
entire set is described as independently prepared; polarizations were taken
from
the literature, or from Kronberg, and axes were fit anew from maps. Again the
data reveals signals with P-values of $10^{-2}$ to $10^{-4}$ contradicting the
null hypothesis.  To be specific, the {\it JM Map 1} correlation is not very
sensitive: Bietenholz's set of 277 points gives $n\rho^2_{1\times 3} = 4.78$, a
weak indication but not statistically significant.  However likelihood analysis
shows that this entire sample is correlated.  The best null distribution is
again the sharply peaked ansatz; the statistic $2T_{Map\ 3} = 8.7$, giving a
$P-value = 3.4\%$.  Axis parameters are gratifyingly in agreement with those of
the other data sets. The strong correlation this set gives for $RM < 0$ region
was already mentioned.  Perfectly consistent, deleting a strip centered on the
mean also gives a big signal in this independent data set.

\subsection{VLA Data} 
\noindent 
There remains a VLA-generated data set of
only
about 30 points created very recently.$^{37}$ Should such data be
included with the set we have analyzed? Unfortunately these data are not
comparable; the set consists of different physical observables.  A close
examination shows that galaxy axes are not used, but selected small pieces of
straight ``jets'' are substituted. Moreover the entire set is taken from a
regime where Faraday rotation is small and generally ignored. Direct
polarizations at high frequencies were then substituted for our polarization
variable $\chi$.  But our $\chi$ comes from actually {\it measuring Faraday
rotation} at different frequencies, and it is physically unjustified to connect
these to the high frequency VLA values. There is unfortunately no
model-independent method to connect the two kinds of data.

Another question that has arisen is one of data ``quality''.  While the VLA
data
are certainly of high quality for the unrelated purpose of seeing fine
structure
inside radio galaxies, there is every reason to believe that the data we use is
of high quality for the statistical purpose we have. We are quite enthusiastic
for the terrific potential of using both kinds of data in complementary ways,
especially since there is at least one physical mechanism$^{17}$
which is frequency dependent and could exploit the differences.
Indeed this effect requires a frequency regime for Faraday rotation not to be
negligible.  Perhaps the large $RM$ cuts  found associated with the
correlations
are related, an interesting question that requires further investigation.

\section{Conclusions}
\noindent
In presenting a study of restricted scope, our conclusions are most crisply
phrased
in a negative sense: {\it  the null hypothesis of uncorrelated isotropy is not
supported. On the basis of significance, it can be ruled out}.  By the
nature of
this study, one is constrained from concluding prematurely what the correlation
found may represent. Under many separate statistical probes, the evidence
against isotropy in the data is significant at $95\%-99.99\%$ (roughly
$2-4\sigma$) confidence levels.

This is not the first such finding, but just one more among a number of studies
accumulating over the years.   While no evidence of systematic bias is
found, we
strongly reiterate the possibility. Yet the persistence of the effect seems to
indicate physical processes outside the framework which has been used to
interpret the data conventionally.  Associated with this behavior are
persistent
axis parameters concordant with the axis parameters found in Nodland and
Ralston,$^1$
and which subsequently have been found to coincide with the CMB dipole
direction.$^{20,21}$  Nevertheless this
is a new field and it would be premature to fix on a physical origin now.  We
therefore postpone more detailed conclusions, and recommend that physical
models
be used to suggest suitable directions of research.  Local effects, while
traditionally held to be under control, can potentially be ruled out with
redshift information.   Resources exist to generate cosmological radio data
sets
with many more points, and the time may be ripe for clever technological
advances that could be revolutionary.  New analysis combined with new data
might
tell us what is causing the effect.

\bigskip 

\nonumsection{Acknowledgements} 
\noindent
We thank Borge Nodland, Hume Feldman, Doug McKay
and G. K. Shukla for useful comments. Supported by DOE
grant number 85 ER40214, the KU General Research Fund, 
the NSF-K*STAR Program under the Kansas Institute for
Theoretical and Computational Science and DAE grant number DAE/PHY/96152.

\nonumsection{References}

\end{document}